\documentclass[12pt]{article}
\usepackage{latexsym,graphicx,epsfig,psfrag,here}
\newcommand{\beq}{\begin{equation}} 
\newcommand{\eeq}{\end{equation}} 
\newcommand{\beqa}{\begin{eqnarray}} 
\newcommand{\eeqa}{\end{eqnarray}} 
\newcommand{\beqan}{\begin{eqnarray*}} 
\newcommand{\eeqan}{\end{eqnarray*}} 
\newcommand{\ba}{\begin{array}} 
\newcommand{\ea}{\end{array}}

\newcommand{\ol}{\overline}

\newcommand{\nn}{\nonumber \\}

\newcommand{\bea}{\begin{eqnarray}} 
\newcommand{\eea}{\end{eqnarray}}

\newcommand{\RE}{\mbox{\rm Re}} 
 
\newcommand{\hepph}[1]{{\tt hep-ph/#1}} 
 
\newcommand{\PL}[3]{{Phys. Lett.} {\bf#1} {(#2)} {#3}} 
\newcommand{\PRL}[3]{{Phys. Rev. Lett.}  {\bf#1} {(#2)} {#3}} 
\newcommand{\PR}[3]{{Phys. Rev.} {\bf#1} {(#2)} {#3}} 
\newcommand{\NP}[3]{{Nucl. Phys.} {\bf#1} {(#2)} {#3}} 
\newcommand{\EPJ}[3]{{Eur. Phys. J.} {\bf#1} {(#2)} {#3}} 
\newcommand{\ZP}[3]{{Z. Phys.} {\bf#1} {(#2)} {#3}} 
\normalsize 
\begin{document} 
\begin{titlepage} 
\begin{flushright} 
UWThPh-2001-15\\ 
April 2001\\ 
\end{flushright} 
\vspace{2.5cm} 
\begin{center} 
{\Large \bf Isospin Violation \\[10pt] and the Magnetic Moment of the Muon$^*$} 
\\[40pt] 
V. Cirigliano, G. Ecker and H. Neufeld  
 
\vspace{1cm} 
Institut f\"ur Theoretische Physik, Universit\"at 
Wien\\ Boltzmanngasse 5, A-1090 Vienna, Austria \\[10pt] 
 
\vfill 
{\bf Abstract} \\ 
\end{center} 
\noindent 
We calculate the leading isospin-violating and electromagnetic
corrections for the decay $\tau^- \to \pi^0 \pi^- \nu_\tau$ at low
energies. The corrections are small but
relevant for the inclusion of $\tau$ decay data in the determination 
of hadronic vacuum polarization especially for the anomalous magnetic
moment of the muon. We show that part of the systematic
differences between the measured form factors in 
$\tau^- \to \pi^0 \pi^- \nu_\tau$ and $e^+ e^- \to \pi^+ \pi^-$
is due to isospin violation.
 
\vfill 
\noindent *~Work supported in part by TMR, EC-Contract  
No. ERBFMRX-CT980169 (EURODA$\Phi$NE).
 
\end{titlepage} 
\addtocounter{page}{1} 
\paragraph{1.}
Precise knowledge of hadronic vacuum polarization is essential for a
reliable determination of both the running of the QED fine structure
constant and of the anomalous magnetic moment of the muon $a_\mu$. For 
the latter, the low-energy structure of hadronic vacuum polarization is
especially important. In fact, about 70 $\%$ of $a_\mu^{\rm vacpol}$
is due to the two-pion intermediate state for $4 M_\pi^2 \le t \le
0.8$ GeV$^2$ (see, e.g., Ref. \cite{narison}). 

A precision of 1 $\%$ has been achieved in the calculation of 
$a_\mu^{\rm vacpol}$ by including \cite{alemany} the more accurate 
$\tau$ decay data \cite{taudata} in addition to experimental results for 
$\sigma(e^+ e^- \to $ hadrons). This is possible because of a CVC relation
between electromagnetic and weak form factors in the
isospin limit. However, both the aforementioned theoretical accuracy 
and the new high-precision experiment at Brookhaven \cite{bnl} warrant 
a closer investigation of isospin violation. A crucial quantity in this
connection is the pion mass difference
$M_{\pi^+}^2-M_{\pi^0}^2=0.067 ~\ol{M_{\pi}^2}$
that is almost exclusively due
to electromagnetic effects. Therefore, both the light
quark mass difference and electromagnetism have to be taken into
account in a consistent treatment of isospin violation.

We concentrate in this letter on isospin 
violation in the 
reactions $\tau^- \to \pi^0 \pi^- \nu_\tau$ and $e^+ e^- \to \pi^+
\pi^-$ at low energies. Chiral perturbation theory (CHPT) \cite{chpt}
is the only framework where such corrections can be reliably
calculated for the standard model in a systematic low-energy
expansion. More specifically, we are going to calculate the
leading corrections of both $O[(m_u-m_d)p^2]$ and $O(e^2 p^2)$ for the 
CVC relation between the two-pion (vector) form factors in the two
processes. A systematic chiral counting will be essential to extract
the leading effects at low energies.

\paragraph{2.} 
The contribution of hadronic vacuum polarization at $O(\alpha^2)$ to
the anomalous magnetic moment of the muon $a_\mu=(g_\mu-2)/2$ is given
by \cite{GdR69}
\begin{equation}
a_\mu^{\rm vacpol}=\displaystyle\frac{1}{4\pi^3}\displaystyle\int_{4
M_\pi^2}^{\infty} dt K(t) \sigma^0_{e^+ e^- \to  {\rm hadrons}}(t)
\end{equation} 
where $K(t)$ is a smooth kernel concentrated at low energies.
The superscript in $\sigma^0_{e^+ e^- \to  {\rm hadrons}}$ denotes the 
``pure'' hadronic cross section with QED corrections removed \cite{ej95}.
For the two-pion final state under discussion this means that $F_V(t)$
in
\begin{eqnarray} 
\sigma^0_{e^+ e^- \to \pi^+ \pi^-}(t)&=&\displaystyle\frac
{\pi\alpha^2\beta^3_{\pi^+\pi^-} (t) }{3 t}|F_V(t)|^2 \\
\beta_{\pi^+\pi^-} (t) &=&\lambda^{1/2}(1,M^2_{\pi^+}/t,M^2_{\pi^+}/t)=
(1-4 M^2_{\pi^+}/t)^{1/2} \nn
\lambda(x,y,z)&=&x^2+y^2+z^2-2(xy+yz+zx)
\end{eqnarray} 
is the vector form factor of the pion with QED turned off (except for
electromagnetic contributions to the charged meson masses).

The decay $\tau^- \to \pi^0 \pi^- \nu_\tau$ is in general governed by
two form factors $f_+,f_-$. In the absence of electromagnetic
corrections, these form factors are functions of the single variable
$t$ that is again the invariant mass squared of the two pions in the
final state. The inclusion of electromagnetic effects generates shifts
to the form factors which depend on a second Dalitz variable 
$u=(P_\tau - p_{\pi^-})^2$. Denoting by $f_+ (t) ,f_- (t)$ the 
$u$-independent components of the form factors (to be defined precisely 
below), 
the decay distribution with respect to $t$ takes the general form
\begin{eqnarray} 
\label{dGamma} 
\displaystyle\frac{d \Gamma(\tau^-\to \pi^0 \pi^- \nu_\tau)}{dt}&=&
\displaystyle\frac{\Gamma_e^{(0)} S_{\rm EW}|V_{ud}|^2}{2 m_\tau^2}
\beta_{\pi^0\pi^-}(t) \, (1-\frac{t}{m_\tau^2})^2  
\left\{|f_+(t)|^2 \left[(1+\frac{2t}{m_\tau^2})\beta^2_{\pi^0\pi^-}(t)
\right.\right.\nn
&&+\left.\left.\frac{3\Delta_\pi^2}{t^2}\right] +  3|f_-(t)|^2
-6 \RE~[f_+^*(t)f_-(t)]\frac{\Delta_\pi}{t}\right\} \, G_{\rm EM} (t) 
\end{eqnarray} 
with
$$
\Gamma_e^{(0)}=\frac{G_F^2 m_\tau^5}{192 \pi^3} ~,~
\Delta_\pi=M_{\pi^+}^2-M_{\pi^0}^2~,~
 \beta_{\pi^0\pi^-} (t) =\lambda^{1/2}(1,M^2_{\pi^0}/t,
M^2_{\pi^+}/t)~.
$$
The factor $S_{\rm EW}$ takes into account the dominant short-distance
electroweak corrections \cite{MS88}.  In the discussion of
semi-leptonic $\tau$ decays, the QED scale of $S_{\rm EW}$ is usually
chosen at the $\tau$ mass.  Thus, to lowest order in $\alpha$, the
short-distance enhancement factor is given by $S_{\rm EW} = 1 +(\alpha
/ \pi) {\rm log}(M_Z^2/m_{\tau}^2)$.  Including the dominant
electromagnetic higher-order effects, one finds the commonly used
value \cite{alemany} $S_{\rm EW} = 1.0194$.  The factor $G_{\rm EM}
(t)$ arises from the integration of the $u$-dependent electromagnetic
correction over the Dalitz variable $u$.  In principle, the spectrum
distortion $G_{\rm EM}(t)$ receives both virtual and real photon
contributions. The measured electronic decay rate of the $\tau$ lepton is 
related to $\Gamma_e^{(0)}$ by \cite{KinSir59}
\beq \label{leprate}
\Gamma(\tau^- \to e^- \bar\nu_e \nu_\tau (\gamma) ) = 
\Gamma_e^{(0)} \bigg[ 1+O(\frac{m_e^2}{m_\tau^2}) \bigg]
\bigg[ 1+\frac{\alpha}{2\pi} (\frac{25}{4} - \pi^2) + O(\alpha^2) \bigg]
~.~
\eeq

In the isospin limit\footnote{From now on, the term isospin limit
stands for both $m_u=m_d$ and $e=0$.} we have $M_{\pi^+}=M_{\pi^0}$,
$S_{\rm EW} = G_{\rm EM} (t) = 1$ and
\begin{eqnarray} 
f_+(t) &=& F_V(t) \nn
f_-(t) &=&0 ~,
\end{eqnarray} 
implying the CVC relation 
\begin{eqnarray} 
\sigma^0_{e^+ e^- \to \pi^+ \pi^-}(t)&=&\displaystyle\frac{1}
{{\cal N}(t) \, \Gamma_e^{(0)}}\displaystyle\frac{d \Gamma(\tau^-\to \pi^0 
\pi^-
\nu_\tau)}{dt} \nn
{\cal N}(t)&=&\displaystyle\frac{3 |V_{ud}|^2}{2 \pi \alpha^2 m_\tau^2} 
t(1-\frac{t}{m_\tau^2})^2 \, (1+\frac{2t}{m_\tau^2})~.
\label{CVC}
\end{eqnarray} 

Including isospin violation to leading order, $O[(m_u-m_d)p^2]$ and 
$O(e^2 p^2)$, we find from Eq.~(\ref{dGamma}) that still only the form
factor $f_+(t)$ survives to this order. The modified CVC relation
takes the form 
\begin{eqnarray} 
\sigma^0_{e^+ e^- \to \pi^+ \pi^-}(t) & = & \displaystyle\frac{1}
{{\cal N}(t) \, \Gamma_e^{(0)}}\displaystyle\frac{d \Gamma(\tau^-\to \pi^0 
\pi^-
\nu_\tau)}{dt} \, \displaystyle\frac{R_{\rm IB} (t)}{S_{\rm EW}} \,  \\
 R_{\rm IB} (t) & = &   
\displaystyle\frac{1}{G_{\rm EM}(t)} \displaystyle\frac{\beta^3_{\pi^+ 
\pi^-}(t)}{\beta^3_{\pi^0\pi^-}(t)} \left|\displaystyle\frac{F_V(t)}{f_+(t)}
\right|^2 ~. \label{riso}
\end{eqnarray}

Bremsstrahlung of soft photons (in principle
contained in the function $G_{\rm EM}(t)$) is subtracted (at least in some
approximation to be discussed below) directly from the raw data 
\cite{taudata}. In the analysis of Ref.~\cite{alemany}, some
additional isospin-violating corrections such as the width difference
$\Gamma_{\rho^+}-\Gamma_{\rho^0}$ were applied. The importance of the
phase space correction factor $\beta^3_{\pi^+ 
\pi^-}(t)/\beta^3_{\pi^0\pi^-}(t)$ has very recently been emphasized
in Ref.~\cite{kuehn}.

It is the purpose of this work to estimate
the remaining contributions to $R_{\rm IB} (t)$.  Working at leading
order, the form factor $F_V(t)$ needs to be calculated to
$O[(m_u-m_d)p^2]$ with physical meson masses (but without explicit
photonic corrections) whereas $f_+(t)$ must be calculated to both
$O[(m_u-m_d)p^2]$ and $O(e^2 p^2)$ if $d\Gamma/dt$ is to be extracted
from actual $\tau$ decay data.

\paragraph{3.} 
To first order in isospin breaking and to first non-trivial order in
the low-energy expansion, isospin violation manifests itself in the
pion vector form factor $F_V(t)$ only in the masses of the particles
contained in the loop amplitude:
\begin{equation} 
F_V(t)=1+2 H_{\pi^+\pi^-}(t) + H_{K^+ K^-}(t)
\label{FVp4}
\end{equation}
with \cite{gl852}
\begin{equation} 
H_{PQ} (t) =\displaystyle\frac{1}{F^2} \bigg[ h_{PQ}(t,\mu) 
+ \frac{2}{3}t L_9^r(\mu) \bigg] ~, 
\end{equation} 
where $F$ denotes the pion decay constant in the chiral limit. The 
expression for the loop function $h_{PQ}(t,\mu)$ is reported in
the Appendix.
The low-energy constant $L_9^r(\mu)$ governs the charge radius of the
pion which is in turn completely dominated by the $\rho$ resonance. 
We will use the prescription of Ref.~\cite{gpgdpp} where the
CHPT form factor (\ref{FVp4}) of $O(p^4)$ was matched to the resonance
region:
\begin{eqnarray} 
F_V(t)&=&\displaystyle\frac{M_\rho^2}{M_\rho^2 - t -i M_\rho
\Gamma_\rho(t)}  \exp{ \bigg[2{\tilde H}_{\pi^+\pi^-}(t)+{\tilde
H}_{K^+ K^-}(t) \bigg]}~,
\label{FVrho}
\end{eqnarray}
with the hadronic off-shell width (for the present case of the
$\rho^0$, the charged pion and kaon masses must be inserted)
\begin{equation} 
\Gamma_\rho(t)=\displaystyle\frac{M_\rho t}{96 \pi F_\pi^2}
\left[\beta^3_{\pi\pi}(t)\theta(t-4 M_\pi^2)+\frac{1}{2}
\beta^3_{KK}(t) \theta(t-4 M_K^2)\right]
\label{width}
\end{equation}
and with a subtracted loop function (setting $\mu=M_\rho$)
\begin{equation} 
{\tilde H}_{PQ}(t)=\displaystyle\frac{\RE~h_{PQ}(t,M_\rho)}{F^2}~.
\end{equation} 
The representation (\ref{FVrho}) has the following attractive features
\cite{gpgdpp}:
\begin{itemize} 
\item By construction, it has the correct low-energy behaviour to
$O(p^4)$ and its asymptotic behaviour is in accordance with QCD;
\item It gives an excellent description of $e^+ e^- \to \pi^+
\pi^-$ data up to $t\sim 1$ GeV$^2$ with the single parameter
$M_\rho\simeq$ 775 MeV.
\end{itemize}
For our present purposes, the representation (\ref{FVrho}) exhibits
in addition the correct behaviour to first order in isospin violation.
To the order we are working, neither the $\rho^+-\rho^0$ mass
difference nor isospin-violating corrections to $F_\pi$ (we use 
$F=F_\pi=92.4$ MeV) enter. All the
isospin violation to this order is contained in the physical meson
masses in $\Gamma_\rho(t)$ and ${\tilde H}_{PQ}(t)$ and this feature
will carry over to the form factor $f_+(t)$, except for additional
purely electromagnetic corrections.

At our level of precision, $\rho-\omega$ mixing does not appear
either. Such higher-order effects are not necessarily negligible
numerically (see, e.g., Ref.~\cite{maltman}).
They can be and partly are taken into
account in the actual analysis of the data (see, e.g., 
Ref.~\cite{alemany}) or can be included in the theoretical error of 
$a_\mu^{\rm vacpol}$.

\paragraph{4.} 
To first order in isospin violation, this time including explicit
photonic corrections, the form factor $f_+(t,u)$ is given by
\begin{equation} 
f_+(t,u)=1 + 2 H_{\pi^0\pi^-}(t) + H_{K^0 K^-}(t)
           +  f_{\rm loop}^{\rm elm}(u,M_\gamma)
+ f_{\rm local}^{\rm elm}~. 
\label{fplus1}
\end{equation}

\begin{figure}
\centerline{\epsfig{file=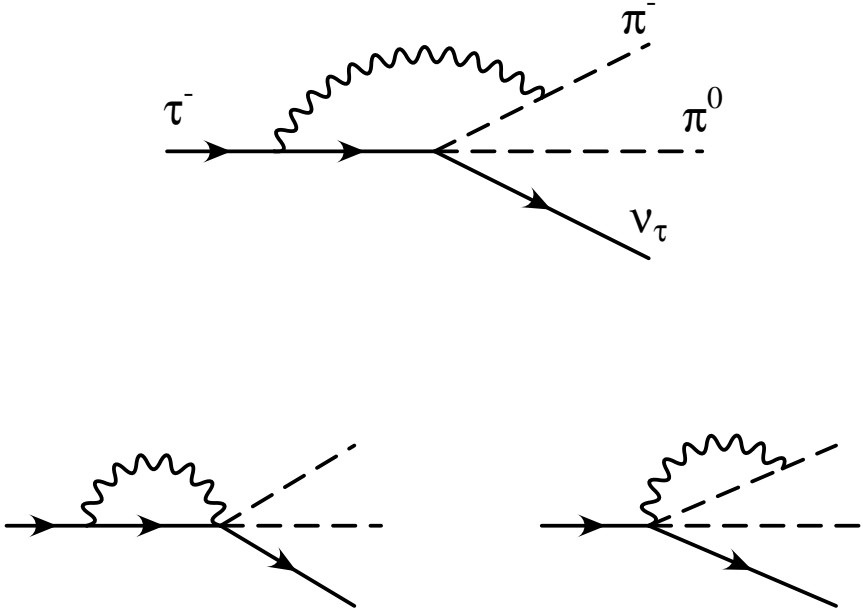,height=8cm}}
\caption{Photon loop diagrams (without wave function renormalization).}
\label{fig:phloop}
\end{figure}
\noindent
Compared to the form factor $F_V(t)$ in (\ref{FVp4}), the appropriate
meson masses appear in the loop amplitude and there is an additional
electromagnetic amplitude containing both the photon loop diagrams
shown in Fig.~\ref{fig:phloop} and an associated local part. The 
electromagnetic amplitude depends on the second Dalitz variable
$u=(P_\tau - p_{\pi^-})^2$ but not on $t$.  Using a small photon mass
$M_\gamma$ as infrared regulator, the electromagnetic amplitudes 
are given by (the
corresponding calculation for $K_{l3}$ will be presented in
Ref.~\cite{kl3} with more details)
\begin{eqnarray}
 f_{\rm loop}^{\rm elm}(u,M_\gamma) & = &  \frac{\alpha}{4 \pi} \bigg[
(u - M_{\pi}^2) \, {\cal A} (u) + 
(u - M_{\pi}^2 - m_{\tau}^2) \,  {\cal B} (u)  \nonumber \\*
& &  +  2 \, (M_{\pi}^2 + m_{\tau}^2 - u) \, {\cal C} (u,M_{\gamma}) 
\bigg]  \label{phloop} 
\end{eqnarray}
\begin{equation}
f_{\rm local}^{\rm elm} = \frac{\alpha}{4 \pi} \bigg[ - \frac{3}{2} 
- \frac{1}{2} \log \frac{m_{\tau}^2}{\mu^2} 
- \log \frac{M_{\pi}^2}{\mu^2} + 2 \log \frac{m_{\tau}^2}{M_{\rho}^2}
- (4 \pi)^2 
\left(- 2 K_{12}^{r} (\mu) + \frac{2}{3} X_1 
+ \frac{1}{2} {\tilde X}_6^{r} (\mu) \right) \bigg] \ .
\end{equation}
Expressions for the functions ${\cal A} (u)$, ${\cal B} (u)$, and 
${\cal C} (u,M_{\gamma})$  are given in the Appendix. We have included
logarithmic terms in $f_{\rm local}^{\rm elm}$ to make it scale
independent. 
The coupling constants $K_{12}$, $X_1$, and $X_6$ appear in the 
low-energy expansion of the standard model at order $e^2 p^2$ 
with inclusion of virtual photons \cite{urech} and of both virtual 
photons and leptons \cite{lept}. Here we have pulled out the 
short-distance part $X_6^{\rm SD}$ of $X_6$ by the 
decomposition \cite{kl3} 
\begin{equation}
X_6^r(\mu) =  X_6^{\rm SD} + 
\tilde{X}_6^r(\mu)
\label{decomp}
\end{equation}
where 
\begin{equation}
e^2 X_6^{\rm SD} = 
- \frac{e^2}{4 {\pi}^2}
 \log \frac{M_Z^2}{M_{\rho}^2} = 1 - S_{\rm EW}
- \frac{e^2}{4 {\pi}^2}
 \log \frac{m_{\tau}^2}{M_{\rho}^2} ~.
\label{X6SD}
\end{equation}
The size of these contributions is discussed in the next section. 

The loop function $f_{\rm loop}^{\rm elm}(u,M_\gamma)$ encodes
universal physics related to the Coulomb interaction between the
$\tau$ and the charged pion. In other words, the $u$-dependence of the
loop amplitude has little to do with low-energy QCD and thus with the
chiral expansion. Rather, it represents the contribution to the loop
integral given by low-energy virtual photons (the ultraviolet part being
absorbed in the definition of the local amplitude). It is therefore
natural to factorize these universal effects in an overall term
\cite{yfs}.  Moreover, since this factorization does not rely on
chiral counting, we are lead to write
(cf. Eq.~(\ref{FVrho})):
\begin{eqnarray} 
f_+(t,u) & = & f_{+} (t) \bigg[1 + f_{\rm loop}^{\rm elm}(u,M_\gamma)
\bigg] \\ 
f_+(t)&=&\displaystyle\frac{M_\rho^2}{M_\rho^2 - t -i M_\rho
\Gamma_\rho(t)}  \exp{ \bigg[2{\tilde H}_{\pi^0\pi^-}(t)+{\tilde
H}_{K^0 K^-}(t) \bigg]} + f_{\rm local}^{\rm elm} ~.
\label{fplus2}
\end{eqnarray}
As in the case of $F_V(t)$ in (\ref{FVrho}), this representation of 
$f_{+} (t)$ has the correct low-energy behaviour to $O(p^4)$ and
it interpolates smoothly to the resonance region.
The resonance width $\Gamma_\rho (t)$ in (\ref{width}) has to
be calculated now with the appropriate $\pi^- \pi^0$ and $K^- K^0$
thresholds and phase space factors.

The photon loop amplitude $f_{\rm loop}^{\rm elm}(u,M_\gamma)$ is 
infrared divergent depending on an artificial photon mass
$M_\gamma$. This dependence is canceled by bremsstrahlung of soft
photons making the decay distribution in $(t,u)$ infrared finite.  The
sum of real and virtual contributions produces the following correction
factor to the $(t,u)$ decay distribution 
\begin{equation}
\Delta (t,u) = 1 + 2 f_{\rm loop}^{\rm
elm}(u,M_\gamma) + g_{\rm brems} (t,u,M_\gamma,E_\gamma^{\rm
min}) ~,
\label{brems}
\end{equation}
which depends on the minimal photon energy $E_\gamma^{\rm min}$
detected in the apparatus and is independent of $M_\gamma$.  This
factor has to be multiplied by the kinematical density $D (t,u)$
(defined in the Appendix) and integrated over the variable $u$ to
produce the term $G_{\rm EM}(t)$ in the decay distribution
(\ref{dGamma}) with respect to $t$:
\begin{equation} 
G_{\rm EM}(t) = \frac{ \displaystyle\int_{u_{\rm min}(t)}^{u_{\rm
 max}(t)} du \, D(t,u) \, \Delta(t,u) }{\displaystyle\int_{u_{\rm
 min}(t)}^{u_{\rm max}(t)} du \, D (t,u)} ~.
\label{gem}
\end{equation}  

The details of soft photon emission (and the function $g_{\rm
brems} (t,u,M_\gamma,E_\gamma^{\rm min})$) depend on the
specific experimental setup.  To the best of our knowledge, all 
$\tau$ decay experiments
relevant here \cite{taudata} apply bremsstrahlung corrections in the
same (approximate) way described in Ref.~\cite{was}: only the leading
term in the Low expansion (proportional to $1/E_\gamma$ in the
amplitude) is taken into account including also the logarithmic term in 
the loop amplitude $f_{\rm loop}^{\rm elm}(u,M_\gamma)$ depending on
$M_\gamma$ (contained in
the function ${\cal C} (u,M_{\gamma})$ given in the Appendix).  
As emphasized in Ref.~\cite{was}, this is only an approximate
treatment of bremsstrahlung that can be trusted for sufficiently
small $E_\gamma^{\rm min}$. 
Assuming this prescription, the setup-independent part of $\Delta (t,u)$
therefore involves only the subtracted loop amplitude ($x$ is 
defined in the Appendix) 
\begin{equation} 
f_{\rm loop,sub}^{\rm elm}(u) = f_{\rm loop}^{\rm elm}(u,M_\gamma) +
 \frac{\alpha}{2 \pi} \, (M_{\pi}^2 + m_{\tau}^2 - u) \,    
\frac{1}{m_\tau M_\pi} \frac{x}{1 - x^2}
\log x \log \frac{M_{\gamma}^2}{m_\tau M_\pi}  
\label{fsub}
\end{equation} 
replacing $f_{\rm loop}^{\rm elm}(u,M_\gamma)$ in (\ref{brems}). 

Following Ref.~\cite{yfs}, we have factored out a universal loop
amplitude $f_{\rm loop,sub}^{\rm elm}(u)$. Although this factorization
is independent of the low-energy expansion it is interesting to analyse
the dependence on the lepton mass $m_\tau$. Unlike in $K_{l3}$ decays
\cite{kl3}, the charged lepton is not light compared to a typical 
hadronic scale $\sim M_\rho$. However, we can perform an
expansion in $p/m_{\tau}$ in complete analogy to heavy baryon CHPT
where $p$ stands for a typical meson mass or momentum. Expanding
$f_{\rm loop,sub}^{\rm elm}(u)$ in inverse powers of $m_\tau$ yields 
\begin{equation} 
f_{\rm loop,sub}^{\rm elm}(u)= \frac{\alpha}{4 \pi} \left( - 1 
+ \log \frac{m_{\tau}^2}{M_{\pi}^2} + O(\frac{p}{m_\tau}) \right)~. 
\label{phloopexp}
\end{equation} 
It turns out that the function $G_{\rm EM}(t)$ is quite insensitive to
whether it is calculated from the full $f_{\rm loop,sub}^{\rm elm}(u)$ 
in (\ref{fsub}) or from  its large-$m_\tau$ approximation 
(\ref{phloopexp}). The difference is negligible in the full range 
$4 M_\pi^2 \le t \le 0.8$ GeV$^2$: the leading term in
(\ref{phloopexp}) provides an excellent approximation.

\paragraph{5.}  

\begin{figure}
\centering
\begin{picture}(300,220)  
\put(-5,70){\makebox(100,120){\epsfig{figure=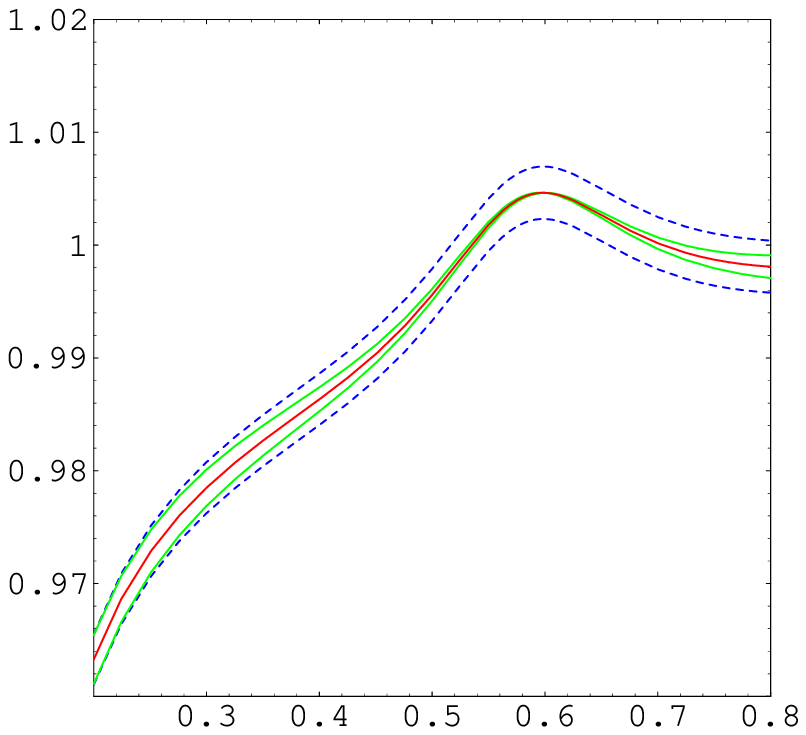,height=6.0cm}}}
\put(120,40){\scriptsize{$t$ (GeV$^2$)}}
\put(-75,200){\scriptsize{$R_{\rm IB} (t)$}}
\put(40,20){\scriptsize{$(a)$}}
\put(220,70){\makebox(100,120){\epsfig{figure=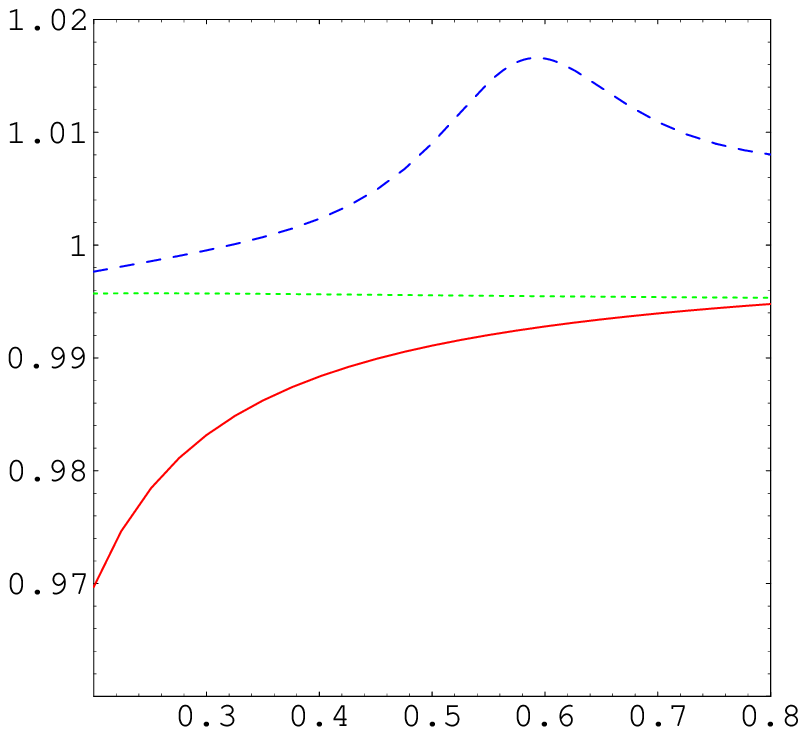,height=6.0cm}}}
\put(340,40){\scriptsize{$t$ (GeV$^2$)}}
\put(280,20){\scriptsize{$(b)$}}
\end{picture}
\caption{$(a)$ Correction factor $R_{\rm IB}(t)$ for isospin violation. 
The bands around the central curve correspond to
the uncertainty in the low-energy constants (solid lines)  
and in the bremsstrahlung factor (dashed lines).  
$(b)$ The separate factors defining  $R_{\rm IB}(t)$ in 
Eq.(\protect\ref{riso}) are plotted as solid line for
$\beta_{+-}^3/\beta_{0-}^3$, dashed line for $|F_V(t)/f_+(t)|^2$ and
dotted line for $1/G_{\rm EM}(t)$.}
\label{fig:RIB}
\end{figure}

The results of our analysis are summarized in Figs.~\ref{fig:RIB}
$(a)$,$(b)$ where we plot the function $R_{\rm IB}(t)$ and its component
factors defined in Eq.~(\ref{riso}) for $ 0.2 \leq t \leq 0.8 $
GeV$^2$.  We note that the dominant contribution at low $t$ is given
by the kinematical term $\beta_{+-}^3/\beta_{0-}^3$
\cite{kuehn}. Photonic corrections embodied in $G_{\rm EM}(t)$ reduce
$R_{\rm IB}(t)$ in addition by about half a percent, largely
independently of $t$. The form factor ratio $|F_V(t)/f_+(t)|^2$ is
dominated by the width difference $\Gamma_{\rho^+}-\Gamma_{\rho^0}$.

We have used the following input for the calculation of
$R_{\rm IB}(t)$.
\begin{itemize}
\item We employ the form factor $F_V (t)$ given in Eq.~(\ref{FVrho}).  
\item We use the form factor $f_+ (t)$ as given in Eq.~(\ref{fplus2}).
The local contribution depends on three low-energy constants appearing
in the chiral expansion. For the constant $K_{12}^{r} (\mu)$ a sum
rule representation is available \cite{moussallam}.  Saturating the
sum rule with low-lying resonances and choosing the QED
renormalization scale between $0.5$ and $1$ GeV, we arrive at the
following estimate: 
$$ K_{12}^{r} (M_\rho) = - (3 \pm 1)\times  10^{-3} ~.$$
As for $X_1$ and $\tilde{X}_6^r (\mu)$,   no estimates are 
presently available. 
We therefore use the upper bounds suggested by  dimensional analysis: 
$$ |X_1| \leq \frac{1}{(4 \pi)^2} ~,~ 
|\tilde{X}_6^r (M_\rho)| \leq \frac{5}{(4 \pi)^2} ~. $$

In the case of  $\tilde{X}_6^r (M_{\rho})$, we have enlarged the naive 
estimate by a factor of 5 (the $\beta$ function associated with $X_6$ 
\cite{lept}). This accounts for the present uncertainty in the matching to 
the short-distance contribution to $X_6$ performed in Eqs.~(\ref{decomp},
\ref{X6SD}). 
The corresponding uncertainty in  $R_{\rm IB} (t)$ is shown in
Fig.~\ref{fig:RIB} $(a)$ (solid curves). 
\item We include the factor $G_{\rm EM}(t)$ according to the discussion
following Eq.~(\ref{brems}).  Due to neglect of sub-leading terms
\cite{was} in the function $g_{\rm brems}
(t,u,M_\gamma,E_\gamma^{\rm min})$, $G_{\rm EM}(t)$ can receive extra
contributions of order $\alpha/(4 \pi) \times O(1)$. We
therefore assign an uncertainty of $\pm \alpha/\pi$ to it.  The effect
on $R_{\rm IB} (t)$ is also shown in Fig.~\ref{fig:RIB}$(a)$ (dashed
curves). Clearly, better knowledge of the radiative amplitude can be
used to improve our determination of $G_{\rm EM} (t)$.

\end{itemize}

In order to quantify the impact of $R_{\rm IB}(t)$ on 
$a_\mu^{\rm vacpol}$, we construct the following ratio: 
\begin{equation}
{\cal R} (t_{\rm max}) = \frac{ \displaystyle\int_{4 M_\pi^2}^{t_{\rm
max}} dt \, K(t) \, \sigma^{0,{\rm CVC}}_{e^+ e^- \to \pi^+ \pi^-  }(t) \, 
R_{\rm IB}
(t)}{ \displaystyle\int_{4 M_\pi^2}^{t_{\rm max}} dt \, K(t) \,
\sigma^{0,{\rm CVC}}_{e^+ e^- \to \pi^+ \pi^- }(t) } ~ , 
\label{ramu}
\end{equation}
where $\sigma^{0,{\rm CVC}}_{e^+ e^- \to \pi^+ \pi^- }(t)$ is obtained 
via CVC from the $\tau$ decay distribution as given in Eq.~(\ref{CVC}). 
A few representative values of ${\cal R} (t_{\rm max})$ are given in 
Table \ref{tab1}. To translate the ratio ${\cal R} (t_{\rm max})$ into
a modification of $a_\mu^{\rm vacpol}(4 M_\pi^2 \le t \le t_{\rm max})$,
we take for the purpose of illustration the value 
$a_\mu^{\rm vacpol}(4 M_\pi^2 \le t \le 0.8~{\rm GeV}^2)= 
(4794.6 \pm 60.7)\times 10^{-11}$ of Ref.~\cite{narison}. Let us assume
for simplicity that this value is calculated from $\tau$ decay data
only. As the quoted number contains the ratio of the radiatively
corrected hadronic rate and the measured electronic mode
(including the radiative channel with a photon in the final state), 
we have to multiply it \cite{Melnikov} by the correction factor 
$1+(\alpha/2\pi)(25/4-\pi^2)$ of Eq.~(\ref{leprate}) in addition to 
${\cal R} (t_{\rm max}= 0.8 \, {\rm GeV}^2)$. In this way, isospin 
violation would reduce $a_\mu^{\rm vacpol}(4 M_\pi^2 \le t \le 
0.8~{\rm GeV}^2)$ by $76 \times 10^{-11}$ or about one standard 
deviation of the reported error.
\renewcommand{\arraystretch}{1.5}
\begin{table}[ht]
\begin{center}
\caption{Correction factor for $a_\mu^{\rm vacpol}$ due to isospin 
violation (defined in 
Eq.~(\protect\ref{ramu})) for some values of $t_{\rm max}$.
An uncertainty of $0.002$ - reflecting the one in the bremsstrahlung 
factor $G_{\rm EM}(t)$ - should be assigned to the values reported here.
This is also an upper bound for the uncertainty due to the low-energy 
constants (see Fig. \ref{fig:RIB}(a)).} 
\label{tab1}
\vspace{.5cm}
\begin{tabular}{|c||c|c|c|}\hline
$t_{\rm max}$ (GeV$^2$) & 0.3 & 0.5 & 0.8\\ \hline
${\cal R} (t_{\rm max})$ & 0.949 & 0.974 & 0.988 \\  
\hline 
\end{tabular}
\end{center}
\end{table}

\paragraph{6.}  
We have calculated the leading isospin-breaking corrections to the CVC
relation between the $e^+ e^- \to \pi^+ \pi^-$ cross section and the
decay distribution for $\tau^- \to \pi^0 \pi^- \nu_\tau$. The
calculation was performed in the framework of CHPT to
$O[(m_u-m_d)p^2)]$ and $O(e^2 p^2)$. Our main result comes in the
form of a function $R_{\rm IB}(t)$ displayed in Fig.~\ref{fig:RIB}$(a)$ 
that
corrects the CVC relation for isospin violation and electromagnetic
effects. Since $R_{\rm IB}(t)$ is smaller than unity in most of the region
under consideration ($4 M_\pi^2 \le t \le 0.8$ GeV$^2$) isospin
violation accounts at least for a sizable part of the systematic
difference at low energies between $e^+ e^-$ and $\tau$ decay data
(e.g., Ref.~\cite{EI99}).

In general, isospin-violating corrections are expected to be of the
order 
\begin{equation} 
\displaystyle\frac{\Delta_\pi}{M_\rho^2}=2 \times 10^{-3}
\label{isoratio}
\end{equation} 
where $M_\rho$ stands for a typical hadronic scale and 
$\Delta_\pi = M_{\pi^+}^2 - M_{\pi^0}^2$. Electromagnetic
corrections embodied in the function $G_{\rm EM}(t)$
are precisely of this magnitude as shown in
Fig.~\ref{fig:RIB}$(b)$. In the form factor
ratio $|F_V(t)/f_+(t)|^2$, the ratio (\ref{isoratio}) is enhanced by 
a numerical factor:
\begin{equation} 
\left|\displaystyle\frac{F_V(M_\rho^2)}{f_+(M_\rho^2)}\right|^2 
\simeq \displaystyle\frac{\Gamma_{\rho^+}^2}{\Gamma_{\rho^0}^2} \simeq
1+ \displaystyle\frac{6 \Delta_\pi}{M_\rho^2-4 M_\pi^2}=1.015 ~.
\end{equation}
The biggest effect occurs in the phase space ratio
$\beta_{+-}^3/\beta_{0-}^3$. It is governed by the function
\begin{equation} 
\displaystyle\frac{3\Delta_\pi}{t- 4 M_\pi^2}
\end{equation} 
and therefore dominates at low energies.
   
Although the calculation is based on a low-energy effective
description of the standard model we claim that the main features of
the correction factor $R_{\rm IB}(t)$ are valid up to $t \simeq 0.8$
GeV$^2$. Of the three factors in the definition (\ref{riso}) of
$R_{\rm IB}(t)$, both the dominant phase space correction factor 
\cite{kuehn}
and the photon loop effects are independent of the low-energy
expansion. Finally, the main part of isospin violation in the form 
factor ratio $|F_V(t)/f_+(t)|^2$ occurs in the $\rho$-width difference 
$\Gamma_{\rho^+}-\Gamma_{\rho^0}$ and should therefore be reliable in
the vicinity of the resonance.

In this work we have been concerned with corrections to the CVC
relation between $\tau$ data and the {\it bare} cross section
$\sigma^0_{e^+ e^- \to \pi^+ \pi^-}$. An important
next-to-leading-order effect of $O(\alpha^3)$ in $a_{\mu}^{\rm had}$ involves 
final state radiative corrections in $\sigma_{e^+ e^- \to \pi^+ \pi^-}$.
Within scalar QED, this was already calculated in Ref.~\cite{schwinger} 
and recently reported in \cite{Melnikov}.
At $O(e^2 p^2)$ in CHPT one obtains the same result \cite{meissner}
as in scalar QED because the local counterterm contributions cancel
due to gauge invariance. The resulting correction to 
$a_{\mu}^{\rm had}$ of $O(\alpha^3)$ is positive \cite{Melnikov}.

\vfill
\paragraph{Acknowledgements}
\noindent 
We thank J. K\"uhn for drawing our attention to the relevance of
isospin violation in the calculation of hadronic vacuum polarization.
We are grateful to J. Gasser, U.-G. Mei\ss ner and K. Melnikov 
for helpful remarks.

\paragraph{Appendix} 

The loop function $h_{PQ} (t,\mu)$ is given by 
\begin{eqnarray} 
h_{PQ} (t,\mu) &=&  \frac{1}{12 t} \lambda (t,M_P^2,M_Q^2) \, 
\bar{J}^{PQ} (t) 
+ \frac{1}{18 (4 \pi)^2} (t - 3 \Sigma_{PQ}) \nn
&& - \frac{1}{12} \left( \frac{2 \Sigma_{PQ} - t}{\Delta_{PQ}} (A_P(\mu) - 
A_Q(\mu))  - 2 (A_P(\mu) + A_Q(\mu)) \right) \ , 
\end{eqnarray} 
where
\begin{eqnarray}
\Sigma_{PQ} & = & M_P^2 + M_Q^2~, \qquad \Delta_{PQ} ~ = ~ M_P^2 -
M_Q^2  \nn 
A_P(\mu)  & = &  - \frac{M_P^2}{(4 \pi)^2} 
\log \frac{M_P^2}{\mu^2}  \nn
\bar{J}^{PQ} (t) & = & \frac{1}{32 \pi^2} \left[ 2 + \frac{\Delta_{PQ}}{t} 
\log \frac{M_Q^2}{M_P^2} - \frac{\Sigma_{PQ}}{\Delta_{PQ}} 
\log \frac{M_Q^2}{M_P^2}  \right. \nn
& & \left.  - \frac{\lambda^{1/2} (t,M_P^2,M_Q^2)}{t} \, 
\log \left( \frac{(t + \lambda^{1/2} (t,M_P^2,M_Q^2))^2 - \Delta_{PQ}^2}{(t - 
\lambda^{1/2} (t,M_P^2,M_Q^2))^2 - \Delta_{PQ}^2} \right) \right]~.
\end{eqnarray}
In terms of the variables 
\begin{equation}
 r_{\tau}  =  \frac{m_{\tau}^2}{M_{\pi}^2} \ , \ \ \ \
 y = 1 + r_{\tau} - \frac{u}{M_{\pi}^2} \ , \ \ \ \    
 x = \frac{1}{2 \sqrt{r_{\tau}}} (y - \sqrt{y^2 - 4 r_{\tau}}) \ ,     
\end{equation} 
and of the dilogarithm 
\begin{equation}
Li_2 (x) = - \int_{0}^{1} \frac{dt}{t} \log (1 - x t)  \ , 
\end{equation}
the functions contributing to $f_{\rm loop}^{\rm elm} (u,M_\gamma)$ 
are given by
\begin{eqnarray}
{\cal A} (u)     &=&   \frac{1}{u} \left[ - \frac{1}{2} \log r_{\tau} + 
\frac{2 - y}{\sqrt{r_{\tau}}} \frac{x}{1 - x^2} \log x \right] \\
{\cal B} (u)     &=&   \frac{1}{u} \left[  \frac{1}{2} \log r_{\tau} + 
\frac{2 r_{\tau} - y}{\sqrt{r_{\tau}}} \frac{x}{1 -
 x^2} \log x \right] \\
{\cal C} (u,M_{\gamma})  &=&   \frac{1}{m_\tau M_\pi} \frac{x}{1 - x^2}
\left[  - \frac{1}{2} \log^2 x + 2 \log x \log (1 - x^2) - 
\frac{\pi^2}{6} + \frac{1}{8} \log^2 r_\tau \right. \nn
 & & \left. + Li_2 (x^2) + Li_2  (1 - \frac{x}{\sqrt{r_\tau}}) + 
Li_2 (1 - x \sqrt{r_\tau}) - \log x \log \frac{M_{\gamma}^2}{m_\tau M_\pi}
 \right] \ . 
\end{eqnarray} 
The kinematical weight to be used in Eq.~(\ref{gem}) is 
\begin{equation}
D (t,u) = \frac{m_\tau^2}{2} (m_\tau^2 - t) + 2 M_{\pi^0}^2 M_{\pi^-}^2  
 - 2 \, u \, (m_\tau^2 - t + M_{\pi^0}^2  +  M_{\pi^-}^2 )  + 2 \, u^2 ~.
\end{equation}
The integration limits in Eq.~(\ref{gem}) are given by 
\begin{eqnarray} 
u_{\rm max/min} (t) & = & a(t) \pm b(t)   \\
a(t) & = & \frac{1}{2} \left[- \Delta_\pi \, \left( 1 + \frac{m_\tau^2}{t}
	\right) +  2 M_{\pi^-}^2 + m_\tau^2 - t \right] \\
b(t) & = & \frac{1}{2}\, \left(m_\tau^2 - t \right) \, \lambda^{1/2} 
	\left(1,\frac{M_{\pi^+}^2}{t},\frac{M_{\pi^0}^2}{t} \right) ~.
\end{eqnarray}


\end{document}